\documentclass[aps,pra,floatfix,twocolumn,showpacs]{revtex4}
\usepackage{epsfig}
\usepackage{graphicx}
\usepackage{dcolumn}
\begin{document}
\title{All-order calculations of the spectra of superheavy elements
  E113 and E114}

\author{T. H. Dinh$^1$ and V. A. Dzuba$^2$}

\affiliation{$^1$Department of Physics, University of Pedagogy, 280 An
  Duong Vuong, Ward 5, Ho Chi Minh City, Vietnam} 

\affiliation{$^2$School of Physics, University of New South Wales,
  Sydney, NSW 2052, Australia} 

\date{\today}

\begin{abstract}
We apply a recently developed method (V. A. Dzuba, PRA {\bf 90},
012517 (2014); J. S. M. Ginges and V. A. Dzuba,  PRA {\bf 91}, 042505
(2015)) to calculate energy levels of superheavy elements Uut
($Z=113$), Fl ($Z=114$), and Fl$^+$. The method combines the
linearized single-double coupled-cluster technigue, the all-order
correlation potential method and configuration interaction
method. Breit and quantum electrodynamic corrections are included. The
role of relativistic and correlation effects is discussed. Similar
calculations for Tl, Pb and Pb$^+$ are used to gauge the accuracy of
the calculations. 

\end{abstract}


\maketitle

\section{Introduction}

The study of the superheavy elements is an important area of research
motivated by the predicted ``island of stability''  
in the region $Z>104$. Elements with nuclear charge up to $Z=118$,
have been synthesised (see, e.g.,
Refs.~\cite{Hamilton13,Turler13,Hess13,gsi,jinr}), and evidence 
for naturally-occurring E122 was reported~\cite{E122}.  
 
Apart from huge activity in the theoretical and experimental nuclear
physics there are also many theoretical works in atomic physics and
quantum chemistry with attempts to predict the chemical prorpeties of
the superheavy elements and their electron structure and spectra (see,
e.g.~\cite{Per,Per15,uzi111-122}). 

Superheavy elements E113 and E114 are of special interest due to their 
closeness to the hypothetical island of stability and relatively
simple electron structure. The E113 atom can be considered as a system
with one external electron above closed-shell core which ends with the
$7s^2$ subshell. Its lighter analog is Tl. The E114 atom can be
considered as a system with two valence electrons. There is a number
of calculations of electron spectra of elements E113 and E114 using
multi-configuration Dirac-Fock, coupled-cluster, configuration
interaction methods and their
combinations~\cite{e13-14a,e13-14b,e13-14c,e13-14d,e13-14e,e13-14f,e13-14g,DF13-14}.
The results of different approaches agree in general trends caused by
interplay of 
relativistic and correlation effects. However, actual numbers for the
energies often differ beyond the uncertainty claimed by the
authors. Therefore, it is important to redo the calculations using the
most advanced techniques which should lead to more accurate and
reliable results. In present paper we apply the recently developed
technique~\cite{Dzuba14} which combines the all-order correlation potential
method~\cite{DFS89}, supplemented by ladder diagrams~\cite{ladder}
with the configuration interaction method~\cite{CI}. The technique
gives very accurate results for energy levels of Cs, Tl, Ba, Lu,
Ra and those ions of these elements which have one or two valence
electrons above closed shells~\cite{ladder,Dzuba14}. It was used to
calculate energy levels of 
superheavy elements E119, E120 and E120$^+$~\cite{E119,Dzuba13,Ginges}. We
demonstrate that the method also works for Pb and Pb$^+$. Then we
apply it to calculate energy levels of E113, E144 and E144$^+$.

\section{Method of calculations}

The method was described in detailed in our previous
papers~\cite{Dzuba14,ladder,Dzuba13,Ginges}.  Here we repeat its main
points with the focus on the details specific for current
calculations. 

\subsection{Atoms with one valence electron}.

Calculations are done in the $V^{N-1}$ approximation which means that
the self-consistent potential is formed by the $N-1$ electrons of the
closed-shell core (the $V^{N-1}$ potential). A complete set of the
single-electron orbitals is obtained by solving the equations
\begin{equation}
h_{0} \psi_{0}=\epsilon_{0} \psi_{0} \ ,
\end{equation}
using the B-spline technique~\cite{Johnson1,Johnson2}.
Here $h_{0}$ is the relativistic Hartree-Fock Hamiltonian
\begin{equation}
\label{hf}
h_{0}=c\mbox{\boldmath$\alpha$}\cdot{\bf p} + (\beta-1)m c^{2} -
\frac{Ze^{2}}{r}+V^{N-1} \ . 
\end{equation}
B-spline basis set and Feynman diagram technique are used to calculate
the all-order correlation potential (CP) $\hat \Sigma$~\cite{DFS89,ladder}.
The CP operator $\hat \Sigma$ is defined in such a way that its expectation value 
for a valence state $v$ is equal to the correlation correction to the energy of this state:
$\delta \epsilon_v = \langle v | \hat \Sigma |v \rangle$.
Perturbation theory expansion for $\hat \Sigma$ starts from the second
order, we use the $\hat \Sigma^{(2)}$ notation for corresponding
CP. Then we include three classes of the higher-order correlations
into the all-order CP $\hat \Sigma^{(\infty)}$~\cite{DFS89}: a)
screening of Coulomb interaction, b) hole-particle interaction, and c)
ladder diagrams~\cite{ladder}. 
States and energies of the valence electron are found by solving the
equation~\cite{corrpot} 
\begin{equation}
(\hat h_{0}+\hat \Sigma) \psi_{v}=\epsilon_{v} \psi_{v} \ .
\label{e:Br}
\end{equation}
Here $\hat \Sigma$ can be either the second-order CP $\hat \Sigma^{(2)}$ or
all-order CP $\hat \Sigma^{(\infty)}$. Note that by iterating the
Eq. (\ref{e:Br}) we include one more class of higher-order
correlations, the iterations of $\hat \Sigma$ (contributions
proportional to $\hat \Sigma^2$, $\hat \Sigma^3$, etc.). The
wave-functions $\psi_v$ of the valence electron found by solving the
Eq.~(\ref{e:Br}) are often called Brueckner orbitals. Corresponding
energies $\epsilon_v$ include correlations. Breit and quantum
electrodynamic corrections are also included (see below).

Table \ref{t:TlPb} presents the results of the calculations for the low
$s$ and $p$ states of Tl and Pb$^+$. The results for Tl are taken from
our earlier paper~\cite{ladder}, the results for Pb$^+$ are obtained
in this work. Contributions of the ladder diagrams are presented
separately because it is the latest addition to the method and it is
important to emphasis its role. Tl and Pb$^+$ have similar electron
structure, therefore it is natural to expect that the results are also
similar. We see however that the results for Pb$^+$ are even slightly
better than for Tl. This is probably due to stronger Coulomb potential
leading to smaller relative value of the correlation
correction. Indeed, the correlation correction to the energy is equal
to the difference between the relativistic Hartree-Fock results (the
RHF column in Table \ref{t:TlPb}) and the experimental values. We see
that the absolute value of the correlation correction is larger for
Pb$^+$ while the relative value is smaller for Pb$^+$ than for Tl. In
the end the accuracy for the energy is on the level of 0.5\%.

\begin{table}
\caption{\label{t:TlPb}
Energy levels of Tl and Pb$^+$ calculated in different
approximations. Final results are the sum of $\Sigma^{\infty}$ and
ladder contributions. $\Delta$ is the difference between final
theoretical results and experimental numbers. The results for Tl are
taken from Ref.~\cite{ladder}. The results for Pb$^+$ obtained in
present work. Experimental numbers are
taken from the NIST database~\cite{NIST}.}
\begin{ruledtabular}
\begin{tabular}{l rrrrrrr}
\multicolumn{1}{c}{State}&
\multicolumn{1}{c}{RHF}&
\multicolumn{1}{c}{$\Sigma^{(2)}$}&
\multicolumn{1}{c}{$\Sigma^{\infty}$}&
\multicolumn{1}{c}{Ladder}&
\multicolumn{1}{c}{Final}&
\multicolumn{1}{c}{$\Delta$}&
\multicolumn{1}{c}{Expt.}\\
\hline
%             RHF       S2       Sall   Ladder  Final Delta  Expt
\multicolumn{8}{c}{Tl}\\
$6p_{1/2}$ & 42823 & 51597 & 50815 & -1215 & 49600 & 336 & 49264\\
$6p_{3/2}$ & 36636 & 43524 & 42491 &  -794 & 41697 & 226  & 41471\\
$7s_{1/2}$ & 21109 & 23375 & 22887 &  -43 &  22844 & 58  & 22786\\
\multicolumn{8}{c}{Pb$^+$}\\

$6p_{1/2}$ & 114360 & 123612 & 122547 & -1421 & 121126 &-119 & 121245\\
$6p_{3/2}$ & 100731 & 109451 & 108108 &  -987 & 107121 &-43  & 107164\\
$7s_{1/2}$ &  58660 &  62793 &  61895 &  -104 &  61791 & -5  & 61796\\

\end{tabular}
\end{ruledtabular}
\end{table}

\subsection{Atoms with two valence electrons}

We use the configuration interaction (CI) technique combined with the
all-order methods to include core-valence
correlations~\cite{Dzuba14,CI,Dzuba13}. The effective CI Hamiltonian
for the system of two valence electrons has the form
\begin{equation}
\hat H^{\rm CI} = \hat h_1(r_1) + \hat h_1(r_2) + \hat h_2(r_1,r_2),
\label{e:CI}
\end{equation}
where $\hat h_1$ is the single-electron operator and $\hat h_2$ is the
two-electron operator. The  $\hat h_1$ operator is the sum of the RHF
operator and the CP $\hat \Sigma_1$
\begin{equation}
\hat h_1 = \hat h_0 + \hat \Sigma_1.
\label{e:h1}
\end{equation}
Here the CP  $\hat \Sigma_1$ is the all-order CP considered in
previous section. We introduce index $1$ to stress that this is a
single-electron operator. 

The $\hat h_2$ operator is the sum of Coulomb interaction and the
correlation operator $\hat \Sigma_2$~\cite{Dzuba14}
\begin{equation}
\hat h_2(r_1,r_2) = \frac{e^2}{|r_1-r_2|} + \hat \Sigma_2(r_1,r_2).
\label{e:h2}
\end{equation}
The  $\hat \Sigma_2$ operator appear due to core-valence correlations
and can be understood as screening of Coulomb interaction between
valence electrons by core electrons. This is also the all-order
operator which comes from solving the single-double (SD)
coupled-cluster equations~\cite{Dzuba14}. Note that solving the SD
equation produce both, single-electron CP $\hat \Sigma_1$ and
two-electron correlation operator $\hat \Sigma_2$. However, for many
atomic systems, including those considered in present work and those
considered previously~\cite{Dzuba14,Ginges}, using the
all-order CP $\hat \Sigma^{(\infty)}$ which was discussed in previous
section, leads to better results than using the SD operator  $\hat
\Sigma_1$. 

Table \ref{t:Pb} shows the results of the calculations for Pb. We
present energies and $g$-factors and compare them to the
experiment. The $g$-factors are useful for identification of the
sates. Comparison with experiment shoes that the accuracy for the
energies is on the level of 1-2\% or better.

\begin{table}
\caption{\label{t:Pb}
Calculated excitation energies ($E$, cm$^{-1}$), and $g$-factors for
lowest states of Pb atom.} 
\begin{ruledtabular}
\begin{tabular}{ll rlrl}
&&\multicolumn{2}{c}{This work}&
\multicolumn{2}{c}{Experiment}\\
\multicolumn{2}{c}{State}&
\multicolumn{1}{c}{$E$}&
\multicolumn{1}{c}{$g$}&
\multicolumn{1}{c}{$E$}&
\multicolumn{1}{c}{$g$}\\
\hline
 $6p^{2}$  & $^1$S$_0$   &     0 &  0.0000 &    0  & 0.0   \\
 $6p^{2}$  & $^3$P$_1$   &  7922 &  1.4999 & 7819  & 1.501 \\
 $6p^{2}$  & $^3$D$_2$   & 10940 &  1.2916 & 10650 & 1.269 \\
 $6p^{2}$  & $^3$D$_2$   & 21924 &  1.2085 & 21458 & 1.230 \\
 $6p^{2}$  & $^1$S$_0$   & 29177 &  0.0000 & 29467 & 0.0   \\
 $7s6p$    & $^1$S$^o_0$ & 35109 &  0.0000 & 34959 & 0.0   \\
 $7s6p$    & $^3$P$^o_1$ & 35536 &  1.3509 & 35287 & 1.349 \\
 $6p7p$    & $^3$D$_1$   & 43236 &  0.6707 & 42919 &       \\
 $6p7p$    & $^1$S$_0$   & 44449 &  0.0000 & 44401 & 0.0   \\
 $6p7p$    & $^3$P$_1$   & 44873 &  1.4690 & 44675 &       \\
 $6p6d$    & $^3$F$^o_2$ & 44986 &  0.7962 & 45443 & 0.798 \\
 $6p7p$    & $^3$D$_2$   & 44997 &  1.1739 & 44809 &       \\
 $6p6d$    & $^1$P$^o_1$ & 46132 &  0.8186 & 46068 & 0.864 \\
 $6p6d$    & $^3$D$^o_2$ & 46162 &  1.2715 & 46061 & 1.247 \\
 $6p6d$    & $^3$F$^o_3$ & 46324 &  1.1184 & 46328 & 1.116 \\
 $7s6p$    & $^3$P$^o_2$ & 48765 &  1.4814 & 48188 & 1.496 \\
 $8s6p$    & $^1$S$^o_0$ & 48784 &  0.0000 & 48726 & 0.0   \\
 $8s6p$    & $^3$P$^o_1$ & 48811 &  1.3238 & 48687 & 1.304 \\
 $7s6p$    & $^1$P$^o_1$ & 49892 &  1.1014 & 49440 & 1.131 \\
 $6p9p$    & $^3$D$_1$   & 51422 &  0.6690 & 51321 &       \\
 $6p9p$    & $^1$S$_0$   & 51683 &  0.0000 & 51786 &       \\
\end{tabular}
\end{ruledtabular}
\end{table}

\subsection{Breit and QED correction}

Since we are considering heavy atoms it is important to include Breit
and quantum electrodynamic (QED) corrections. 

The Breit operator in the zero energy transfer approximation has the form:
\begin{equation}
h^{B}=-\frac{\mbox{\boldmath$\alpha$}_{1}\cdot \mbox{\boldmath$\alpha$}_{2}+
(\mbox{\boldmath$\alpha$}_{1}\cdot {\bf n})
(\mbox{\boldmath$\alpha$}_{2}\cdot {\bf n})}{2r} \ ,
\end{equation}  
where ${\bf r}={\bf n}r$, $r$ is the distance between electrons, and
$\mbox{\boldmath$\alpha$}$ is the Dirac matrix. 

We use the radiative potential method introduced in Ref.~\cite{radpot}
to include QED corrections to the energies.  
The radiative potential has the form
\begin{equation}
V_{\rm rad}(r)=V_U(r)+V_g(r)+ V_e(r) \ ,
\label{e:rad}
\end{equation}
where $V_U$ is the Uehling potential and $V_g$ is the potential
arising from the magnetic formfactor, and $V_e$ is the potential
arising from the electric formfactor. 

Both, Breit and QED operators are included to the Hartree-Fock
iterations so that an important relaxation effect is taken into
account~\cite{relax,Breit1,Ginges2}. 
 
\section{Results and discussion}

\begin{table}
\caption{\label{t:E113}
Energy levels (in cm$^{-1}$) of superheavy elements Uut ($Z=113$) and Fl$^+$
($Z=114$) calculated in different approximations. Notations like in
Table~\ref{t:TlPb}.} 
\begin{ruledtabular}
\begin{tabular}{l rrrrrr}
\multicolumn{1}{c}{State}&
\multicolumn{1}{c}{RHF}&
\multicolumn{1}{c}{$\Sigma^{(2)}$}&
\multicolumn{1}{c}{$\Sigma^{(\infty)}$}&
\multicolumn{1}{c}{Ladder}&
\multicolumn{1}{c}{Final}&
\multicolumn{1}{c}{Other}\footnotemark[1]\\
\hline
%             RHF       S2       Sall   Ladder  Final Delta  Expt
\multicolumn{6}{c}{Uut}\\
$7p_{1/2}$ & 54901 & 61929 & 61953 & -2183 & 59770 & 60154 \\
$7p_{3/2}$ & 31557 & 38498 & 36623 &  -497 & 36126 & 34938 \\
$8s_{1/2}$ & 22193 & 24653 & 23761 &  -32 &  23729 & 21313\\

\multicolumn{6}{c}{Fl$^+$}\\

$7p_{1/2}$ & 130420 & 138110 & 138105 & -2333 & 135772 & 137710 \\
$7p_{3/2}$ &  89802 &  99170 &  96708 &  -667 & 96041 & 97329 \\
$8s_{1/2}$ &  60844 &  65316 &  63832 &  -82 &  63750 & 63964 \\

\end{tabular}
\footnotetext[1]{Reference~\cite{DF13-14}}
\end{ruledtabular}
\end{table}

Table \ref{t:E113} shows the results of calculations for Uut (E113)
and Fl$^+$ superheavy elements in the same form as in Table
\ref{t:TlPb} for Tl and Pb$^+$. Comparison shows some interesting
trends. The total value of the correlation correction for superheavy
elements and their lighter analogs are similar but slightly smaller
for the superheavy elements. This is probably due to relativistic
relaxation which leads to increased energy interval between core and
valence states. On the other hand, the contribution of ladder diagrams
is larger for the ground states of E113 and Fl$^+$ than for Tl and
Pb$^+$. Ladder diagrams describe residual Coulomb interaction between
valence electron and the core. Larger contribution probably reflects
the fact that due to relativistic relaxation the superheavy elements
in the ground state have smaller size than their lighter analogs. Since
the total value of the correlation correction to the energies is very
similar for heavy and lighter elements we expect that the accuracy of
the calculations is also very similar, i.e. $\sim 0.5\%$.

The results of present calculations are in a reasonable agreement with
previous SD+CI calculations of Ref.~\cite{DF13-14} (see
Table~\ref{t:E113}). However, they are closer to the results of
coupled-cluster calculations of Ref.~\cite{e13-14a,e13-14b}. This is
true for both, ionization potential and excitation energies. 

\begin{table}
\caption{\label{t:E114}
Calculated excitation energies ($E$, cm$^{-1}$), and $g$-factors for
lowest states of superheavy element Fl.} 
\begin{ruledtabular}
\begin{tabular}{ll rl rr}
&&\multicolumn{2}{c}{This work}&
\multicolumn{2}{c}{Other}\\
\multicolumn{2}{c}{State}&
\multicolumn{1}{c}{$E$}&
\multicolumn{1}{c}{$g$}&
\multicolumn{1}{c}{$E$\footnotemark[1]}&
\multicolumn{1}{c}{$E$\footnotemark[2]}\\
\hline
 $7p^{2}$  & $^1$S$_0$   &     0 &  0.0000 &     0 &     0 \\
 $7p^{2}$  & $^3$P$_1$   & 26780 &  1.4995 & 27316 & 26342 \\
 $7p^{2}$  & $^3$D$_2$   & 29462 &  1.1966 & 29149 & 28983 \\
 $8s7p$    & $^1$S$^o_0$ & 43573 &  0.0000 & 44036 & 43111 \\
 $8s7p$    & $^3$P$^o_1$ & 43876 &  1.3413 & 44362 & 43441 \\
 $7p8p$    & $^3$D$_1$   & 51646 &  0.6670 & 51834 & 51302 \\
 $7p8p$    & $^1$S$_0$   & 52724 &  0.0000 & 53149 & 52487 \\
 $7p8p$    & $^3$P$_1$   & 54842 &  1.4932 & 55414 & 54647 \\
 $7p8p$    & $^3$D$_2$   & 55015 &  1.1713 & 55191 & 54814 \\
 $7p7d$    & $^3$D$^o_2$ & 55814 &  1.1780 & 56988 &  \\
 $7p7d$    & $^1$D$^o_2$ & 55828 &  0.8730 & 57413 &  \\
 $7p7d$    & $^3$F$^o_3$ & 55890 &  1.1138 & 57481 &  \\
 $7p7d$    & $^1$P$^o_1$ & 55910 &  0.8259 & 57244 &  \\
 $9s7p$    & $^1$S$^o_0$ & 57607 &  0.0000 & 57367 &  \\
 $9s7p$    & $^3$P$^o_1$ & 57663 &  1.3316 &  &  \\
 $7p9p$    & $^3$D$_1$   & 60198 &  0.6669 &  & \\
 $7p9p$    & $^1$S$_0$   & 60324 &  0.0000 &  &  \\
 $7p9p$    & $^3$D$_2$   & 61272 &  1.1769 & 57413 &  \\
 $7p8d$    & $^3$F$^o_2$ & 61612 &  0.7717 &  &  \\
 $7p6f$    & $^1$D$_2$   & 61620 &  0.9097 &  &  \\
 $7p6f$    & $^3$G$_3$   & 61650 &  0.8357 & 60291 &  \\
 $7p6f$    & $^3$F$_3$   & 61653 &  1.1917 & 60298 &  \\
 $7p6f$    & $^3$G$_4$   & 61655 &  1.0838 & 60311 &  \\
\end{tabular}
\footnotetext[1]{Reference~\cite{DF13-14}}
\footnotetext[2]{Reference~\cite{e13-14f}}
\end{ruledtabular}
\end{table}

The results for Fl (E114) are presented in Table~\ref{t:E114} and
compared with previous calculations of
Refs.~\cite{DF13-14,e13-14f}. In most of the cases the results of
present work are in between the two earlier results. However, the
difference between all three sets of results is small, $\sim 1\%$. This
is consistent with the estimate of accuracy based on similar
calculations for Pb (see previous section). 

\section{Conclusion}

We apply a recently developed advanced method of atomic structure
calculation which combines three different all-order techniques to
calculate energy levels of superheavy elements E113, Fl and Fl$^+$
with the accuracy  $\sim 1\%$. This represents some improvement to
previous calculations and contributes to the reliability of the
theoretical predictions of the spectra of superheavy elements.

\acknowledgments

This research is funded by Vietnam National Foundation for Science and
Technology Development (NAFOSTED) under grant number
``103.01-2013.38'' and by the Australian Research Council.

\end{document}